# Surface Deformation of New York City from Multitemporal Interferometric Analysis of Sentinel-1 SAR Datasets

**Manoochehr Shirzaei**[1,2]
[1]Department of Geosciences, Virginia Tech, Blacksburg, VA, USA
[2]Virginia Tech National Security Institute, , Blacksburg, VA, USA
Correspondance: shirzaei@vt.edu

**ABSTRACT** Sentinel-1, the Synthetic Aperture Radar (SAR) mission operated by the European Space Agency (ESA) as part of the Copernicus program, provides Free and Open Data with global coverage, transforming our ability to monitor subtle changes in land surface elevation due to natural and anthropogenic process. Here, we use SAR datasets over New York spanning 2015/03/12 to 2022/08/26 and apply an advanced multitemporal SAR interferometric algorithm to measure rates and time series of surface deformation at an unprecedented resolution and accuracy. We found Line-Of-Sight (LOS) subsidence and uplift rates of up to 25.7 $\pm$ 0.2 mm/yr and 8.7 $\pm$ 0.3 mm/yr affecting parts of the city. The LOS velocities, standard deviations, and local incidence angles for ~1,792,000 pixels at ~25m $\times$ 25 m resolution are provided in an online repository.

## I. INTRODUCTION

   Land subsidence poses a significant hazard in New York, particularly in areas with high urbanization and intensive groundwater extraction [1]. The region's geology plays a crucial role, as certain types of soils, like clay and silt, are more susceptible to compression and settling over time.
   Land subsidence in New York can also be influenced by a geological phenomenon known as Glacial Isostatic Adjustment (GIA) [2]. During the last ice age, massive glaciers covered parts of North America, including the New York region. The immense weight of these glaciers caused the Earth's crust to deform and sink slowly over time. As the glaciers receded and melted, the land rebounded, but this process is still ongoing and occurs much slower than the initial deformation.
   GIA contributes to the overall subsidence hazard in New York by adding a layer of complexity to the region's geological setting. The glacial rebound and other factors like groundwater extraction and urbanization can lead to differential subsidence, where certain areas sink at different rates than others [3]. This non-uniform subsidence can lead to uneven settlement and strain on infrastructure, potentially causing damage to roads, bridges, and buildings [4, 5].
   Understanding the interplay between GIA and other factors contributing to land subsidence is crucial for accurately assessing the risks and implementing effective mitigation strategies [3]. Continued monitoring of ground movement, coupled with advanced geotechnical studies, will aid in better predicting and managing the subsidence hazard in New York and ensuring the city's long-term stability and safety [6].
   Interferometric Synthetic Aperture Radar (InSAR) is an advanced remote sensing technique that has proven valuable for monitoring and studying land subsidence [7]. InSAR uses satellite-mounted radar sensors to measure ground deformation over large areas accurately.

InSAR has several advantages for land subsidence monitoring. It provides a comprehensive and cost-effective solution for detecting and measuring ground deformation over large areas, making it ideal for urban environments like New York. Additionally, it can capture subsidence that may occur gradually over months or years, allowing for early detection and proactive measures to mitigate potential hazards. InSAR data can be integrated with other geospatial information to better understand the underlying causes and potential impacts of land subsidence, thereby assisting urban planners and decision-makers in developing appropriate strategies for sustainable development and infrastructure management.

## II. Methods and Datasets

*A. Study area*

New York is a state located in the northeastern region of the United States. The most populous city in the United States and the heart of the state, New York City is a global financial, cultural, and media capital. It is famous for its iconic landmarks, including the Statue of Liberty, Times Square, Central Park, the Empire State Building, Broadway theaters, and numerous museums. The state capital of New York, located in the eastern part of the state, along the Hudson River.

New York City's geology has played a significant role in shaping its landscape and providing important natural resources for its development. The bedrock beneath New York City consists mainly of ancient metamorphic rocks formed over 1 billion years ago. The predominant rock types include gneiss, schist, and marble. These rocks have been subjected to intense heat and pressure over geological time, resulting in their current appearance.

During the last Ice Age, approximately 10,000 to 20,000 years ago, much of North America, including New York City, was covered by massive ice sheets. The movement of glaciers shaped the region's topography, creating valleys and hills. The scraping action of glaciers also led to the formation of Long Island and Staten Island.

New York City's harbor and surrounding estuaries, including the East River and Long Island Sound, are products of rising sea levels after the Ice Age. These water bodies have shaped the city's waterfront and have significantly influenced its development as a major port and trade center.
The eastern part of New York City, including parts of Queens and Brooklyn, lies on a coastal plain. This region is characterized by relatively flat terrain and is composed of unconsolidated sediments, such as sand, silt, and clay, deposited by rivers and the sea.

Beneath the surface of New York City, there are several underground layers of permeable rock and sand known as aquifers. These aquifers are essential sources of groundwater for the city's water supply.

Sea level rise is a significant concern for New York City due to its coastal location and low-lying geography. As global temperatures continue to rise, the melting of polar ice caps and glaciers and the thermal expansion of seawater contribute to the increase in sea levels worldwide. By the end of the 21st century, sea level rise in New York City is projected to be higher than the global average, potentially reaching 1.5 to 6 feet (about 0.46 to 1.83 meters) or more.

Due to its geography, with much of the city built on low-lying coastal areas and islands, New York City is particularly vulnerable to the impacts of rising sea levels. Sea level rise increases the risk of more frequent and severe coastal flooding, especially during storms and high tides.

*B. Datasets*

The SAR data sets spanning from 2015/03/12 to 2022/08/26 include 183 images in ascending (incidence angle = ~38.9°, heading angle = ~347°) orbits of path 33 of the Sentinel-1 C-band satellite.

*C. Multitemporal InSAR algorithm for LOS time series*

We employ a multitemporal SAR interferometric approach to map the surface deformation time series using the algorithm - Wavelet-Based InSAR (WabInSAR) [8]. We apply multi-looking factors of 12 in range and 2 in azimuth directions, resulting in a pixel size of ~25 m x ~25 m. Then, we perform image co-registration using satellite precise ephemeris data and a Shuttle Radar Topography Mission (SRTM) Digital Elevation Model (DEM) and apply the Enhanced Spectral Diversity algorithm [9]. Next, considering thresholds of 150 m for perpendicular and 400 days for temporal baselines, we create 892 interferograms. We remove the flat Earth and topographic effects using a 3-arc-second (90 m) SRTM DEM [10] and satellite precise ephemeris data [11]. After identifying less noisy pixels (elite pixels), we unwrap the interferogram phases using a 2D-minimum-cost-flow algorithm [12] suitable for sparsely distributed pixels [13]. After correcting unwrapped interferograms for the atmospheric delay [14, 15] and orbital errors [16], we apply the reweighted least square method to obtain the Line of Sight (LOS) displacement time series and the velocity of each pixel [17]. We use the vertical rate of the NYBK ( http://geodesy.unr.edu/NGLStationPages/stations/NYBK.sta) Global Navigation Satellite System (GNSS) to calibrate LOS velocity maps and align them with the IGS14 reference frame.

## III. RESULTS

Figure 1A shows the map of LOS velocity over New York City. The overall map is characterized by subsidence primarily due to the GIA effect as indicated in the LOS cumulative distribution function (Fig. 1B). Also, the standard deviation of the LOS velocity is shown in Figure 1C, with most values below 0.1 mm/yr. Most pixels are affected by subsidence, with a median value of -1.5 mm/yr in the LOS direction.

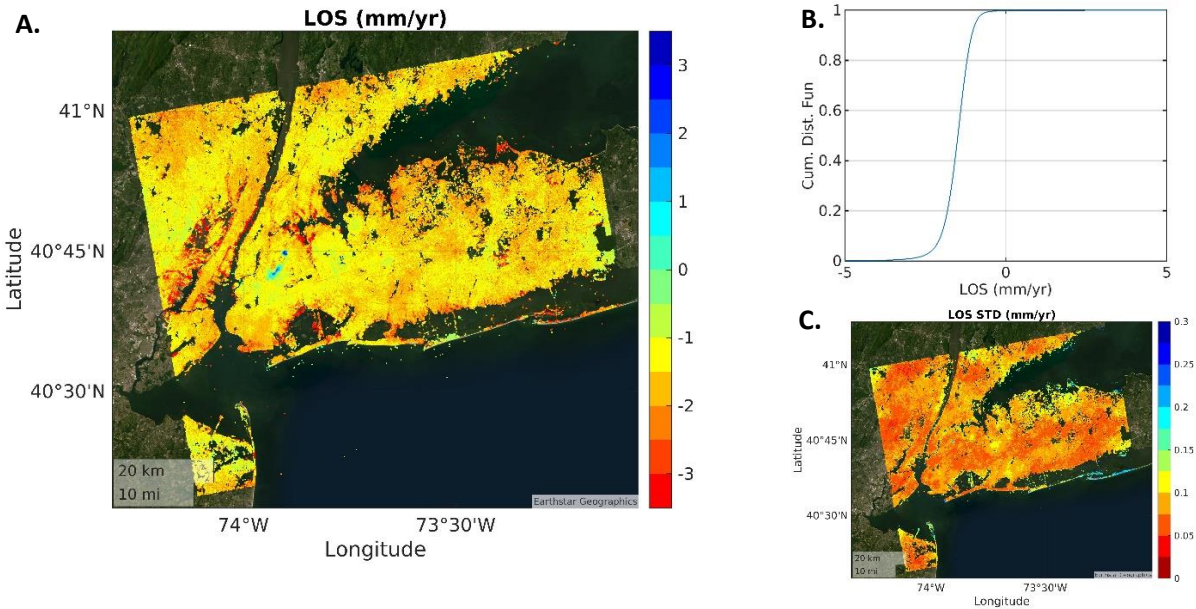

**Figure 1.** A. LOS velocity map including GIA effect. B. Cumulative distribution function of the LOS velocity map. C. Standard deviation of the velocity map.

Next, we remove the GIA effect using an existing model to investigate the non-GIA components of the deformation field [2]. Figure 2A shows the residual map following GIA correction. The long-wavelength subsidence observed in Figure 1 is no longer present, and the cumulative distribution function of LOS has now shifted to the right. In Figure 2A, localized zones of uplift and subsidence are more visible, likely due to anthropogenic and erosional processes.

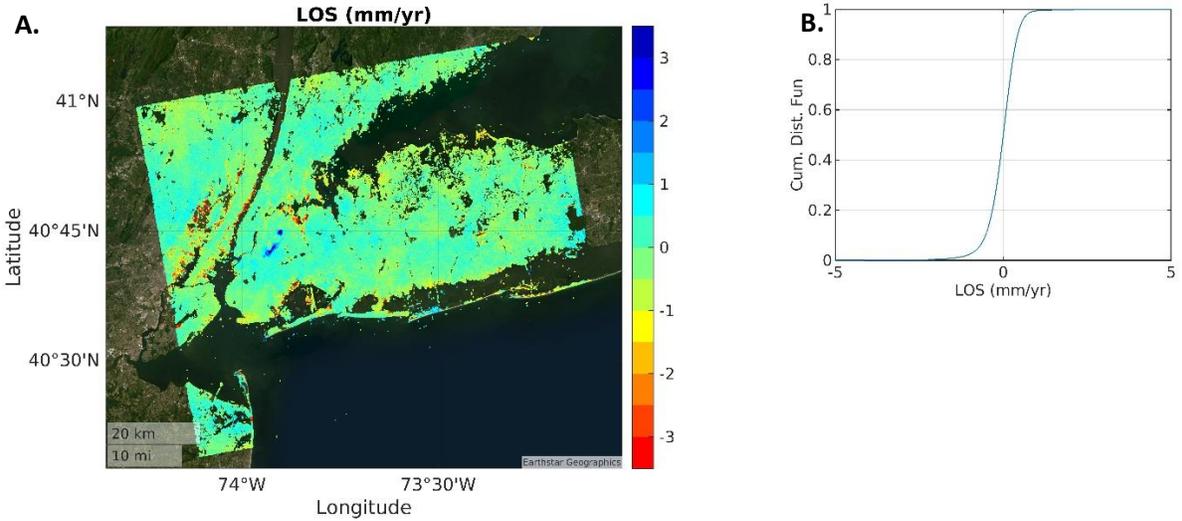

**Figure 2.** A. LOS velocity map after removing GIA effect. B. Emrpical cumulative distribution function of LOS deformation map shown in panel A.

Figure 3A shows a closeup of the uplift area near Longitude -73.91° and latitude 40.74° and associated time series. This is likely the groundwater recharge site, causing increased pore pressure and ground surface rebound.

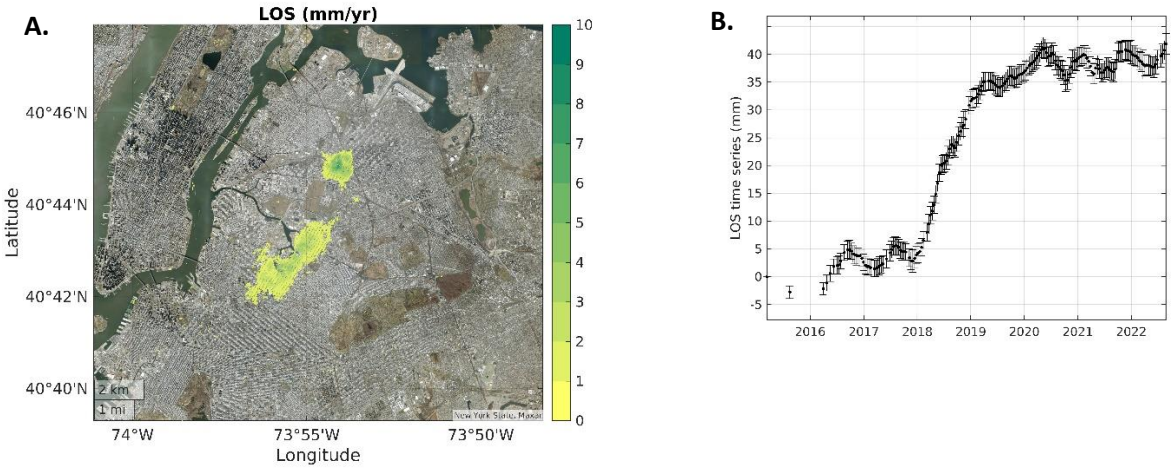

**Figure 3.** A. Uplift signal closeup and example of cumulative LOS time series (B). Errorbars in B are 1sigma errors.

We show the closeup of some subsiding areas and example time series in Figure 4. Most subsidence sites are affected by groundwater depletion and erosional processes.

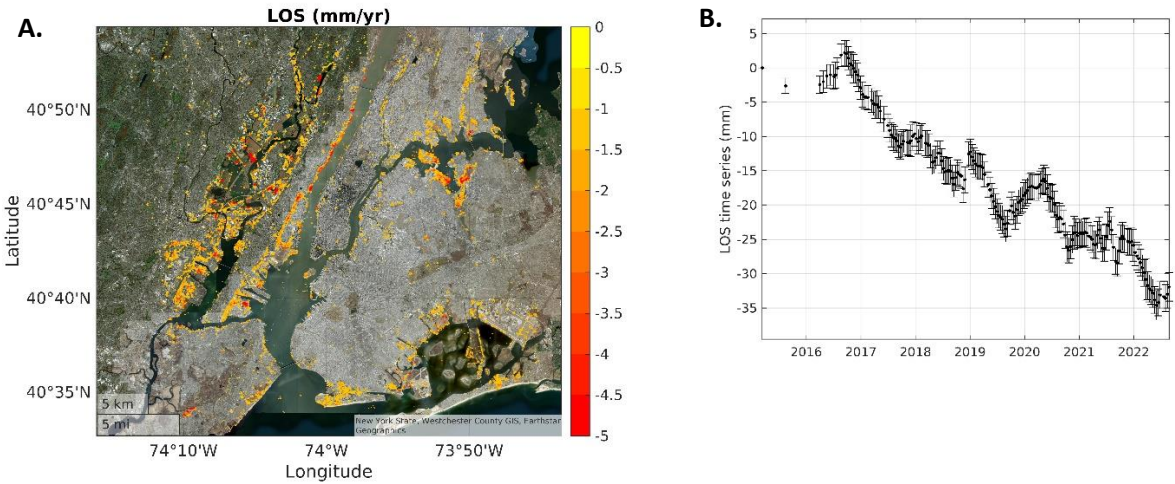

**Figure 4.** A. Subsidence signal closeup and example of cumulative LOS time series (B). Errorbars in B are 1sigma errors.

## IV. OPEN SCIENCE

*A. Data format*

The online repository [18] comprises a CSV file (NY_LLIVS.csv) of 1792308 × 5 dimensions with the following format:
- Column #1: longitude (deg)
- Column #2: Latitude (deg)
- Column #3: Incidence angle (deg)
- Column #4: LOS rate (cm/yr)
- Column #5: LOS rate STD (cm/yr)

*B. Data availability*

Line-of-Sight surface deformation rates at New York from Sentinel-1 SAR interferometric time series analysis are available at https://doi.org/10.7294/23741097

Citation: Shirzaei, Manoochehr (2023). Line-of-Sight surface deformation rate at New York from Sentinel-1 SAR interferometric time series analysis. University Libraries, Virginia Tech. Dataset. https://doi.org/10.7294/23741097

## ACKNOWLEDGMENT

Grants from NASA, USGS, and NSF supported this research. We thank ChatGPT of OpenAI for assistance with developing the introduction.

# REFERENCES

[1] T. Parsons, P. C. Wu, M. Wei, and S. D'Hondt, "The Weight of New York City: Possible Contributions to Subsidence From Anthropogenic Sources," *Earth's Future,* vol. 11, no. 5, p. e2022EF003465, 2023.

[2] W. Peltier, D. Argus, and R. Drummond, "Space geodesy constrains ice age terminal deglaciation: The global ICE-6G_C (VM5a) model," *Journal of Geophysical Research: Solid Earth,* vol. 120, no. 1, pp. 450-487, 2015.

[3] M. Shirzaei, J. Freymueller, T. E. Törnqvist, D. L. Galloway, T. Dura, and P. S. J. Minderhoud, "Measuring, modelling and projecting coastal land subsidence," *Nature Reviews Earth & Environment,* 2020/12/10 2021, doi: 10.1038/s43017-020-00115-x.

[4] L. Ohenhen and M. Shirzaei, "Land Subsidence Hazard and Building Collapse Risk in the Coastal City of Lagos, West Africa," *Earth's Future,* vol. 10, no. 12, p. e2022EF003219, 2022, doi: https://doi.org/10.1029/2022EF003219.

[5] F. Cigna and D. Tapete, "Present-day land subsidence rates, surface faulting hazard and risk in Mexico City with 2014–2020 Sentinel-1 IW InSAR," *Remote Sensing of Environment,* vol. 253, p. 112161, 2021.

[6] P. C. Wu, M. Wei, and S. D'Hondt, "Subsidence in coastal cities throughout the world observed by InSAR," *Geophysical Research Letters,* vol. 49, no. 7, p. e2022GL098477, 2022.

[7] R. Bürgmann, P. A. Rosen, and E. J. Fielding, "Synthetic aperture radar interferometry to measure earth's surface topography and its deformation," *Ann. Rev. Earth Planet. Sci.,* vol. 28, pp. 169-209, 2000.

[8] !!! INVALID CITATION !!! [8-11].

[9] M. Shirzaei, R. Bürgmann, and E. J. Fielding, "Applicability of Sentinel-1 Terrain Observation by Progressive Scans multitemporal interferometry for monitoring slow ground motions in the San Francisco Bay Area," *Geophysical Research Letters,* vol. 44, no. 6, pp. 2733-2742, 2017, doi: 10.1002/2017GL072663.

[10] T. G. Farr, P. A. Rosen, E. Caro, R. Crippen, R. Duren, S. Hensley, M. Kobrick, M. Paller, E. Rodriguez, L. Roth, D. Seal, S. Shaffer, J. Shimada, J. Umland, M. Werner, M. Oskin, D. Burbank, and D. Alsdorf, "The shuttle radar topography mission," *Reviews of Geophysics,* vol. 45, no. 2, May 19 2007, Art no. Rg2004, doi: 10.1029/2005rg000183.

[11] G. Franceschetti and R. Lanari, *Synthetic aperture radar processing*. CRC Press, 1999.

[12] M. Costantini, "A novel phase unwrapping method based on network programming," *Geoscience and Remote Sensing, IEEE Transactions on,* vol. 36, no. 3, pp. 813 - 821, 1998.

[13] M. Costantini and P. A. Rosen, "A generalized phase unwrapping approach for sparse data," in *in Proceedings of the IEEE 1999 International Geoscience and Remote Sensing Symposium (IGARSS)*, Hmburg, 1999, vol. 1, pp. 267-269

[14] J.-C. Lee and M. Shirzaei, "Novel algorithms for pair and pixel selection and atmospheric error correction in multitemporal InSAR," *Remote Sensing of Environment,* vol. 286, p. 113447, 2023, doi: https://doi.org/10.1016/j.rse.2022.113447.

[15] M. Shirzaei and R. Bürgmann, "Topography correlated atmospheric delay correction in radar interferometry using wavelet transforms," *Geophysical Research Letters,* vol. 39, no. 1, p. doi: 10.1029/2011GL049971, Jan 6 2012. [Online]. Available: <Go to ISI>://WOS:000298930900002

[16] M. Shirzaei and T. R. Walter, "Estimating the Effect of Satellite Orbital Error Using Wavelet-Based Robust Regression Applied to InSAR Deformation Data," (in English), *Ieee Transactions on Geoscience and Remote Sensing,* vol. 49, no. 11, pp. 4600-4605, Nov 2011, doi: Doi 10.1109/Tgrs.2011.2143419.

[17] M. Shirzaei, "A Wavelet-Based Multitemporal DInSAR Algorithm for Monitoring Ground Surface Motion," (in English), *Ieee Geoscience and Remote Sensing Letters,* vol. 10, no. 3, pp. 456-460, May 2013, doi: Doi 10.1109/Lgrs.2012.2208935.

[18] M. Shirzaei, "Line-of-Sight surface deformation rate at New York from Sentinel-1 SAR interferometric time series analysis," *University Libraries,* 2023, doi: https://doi.org/10.7294/23741097.